\documentclass[11pt,reqno,a4paper,nofootinbib,aps,epsfig,graphicsfloatfix,mathbbm,pra]{article}
\usepackage{amsmath,amsthm,amsfonts,amssymb,graphics,graphicx,epsfig,color,times,bbm}
\newtheorem{theorem}{Theorem}

\newtheorem{remark}{Remark}

\begin{document}
\title{Exact treatment of operator difference equations with nonconstant and noncommutative coefficients
}
\author{\normalsize M. A.
Jivulescu$^{1}$,  A. Messina$^2$\\
\small ${}^1$  Department of
Mathematics, " Politehnica" University of Timi\c{s}oara,
\\\small P-ta Victoriei Nr. 2,
 300006 Timi\c{s}oara,
Romania\\
\small ${}^2$ MIUR, CNISM and Dipartimento di Fisica, \\\small Universit\`{a} di Palermo, via Archirafi
36, 90123 Palermo, Italy\\
\small {\it Email Address:$^{1}$ maria.jivulescu@mat.upt.ro}\\
\small {\it Email Address:$^2$ messina@fisica.unipa.it}}

\date{}
\maketitle

\begin{abstract}
We study a homogeneous linear second-order difference equation with nonconstant and noncommuting operator coefficients in a vector space. We build its exact resolutive formula consisting of the explicit noniterative expression of a generic term of the unknown sequence of vectors. Some non-trivial applications are reported in order to show the usefulness and the broad applicability of the result.
\end{abstract}

\section{Introduction}
Difference equations naturally occur and play a central role whenever the problem under scrutiny or the phenomenon under investigation allows a mathematical formulation traceable back to the set of natural numbers.
A broad variety of situations in biology, economics, dynamical systems, electrical circuit analysis and other fields \cite[chap 1]{kelley} is indeed modeled by difference equations often demanding, however, quite different resolution methods due to peculiar aspects of their mathematical nature. Thus, like in the case of the differential equations, a useful classification of the difference equations has emerged too.
\\The classical field of difference equations deals with linear or nonlinear equations where
the unknown is a real or complex-valued function defined on a countable domain which in turn  may be thought of
as coincident with $\mathbb{N}^*$ without loss of generality. These equations are conveniently classified as
scalar difference equations \cite[chap 2,3,4]{driver}-\cite{napoli} to discriminate them from the so-called matrix difference equations \cite{taher}, \cite{Jivulescu2} where the unknowns constitute a sequence of $d\times d$ matrices with entries in $\mathbb{C}$ as well as possibly all the given coefficients.

 Quite recently a new class of difference equations has been introduced, namely the class of the linear operator difference equations \cite{Jivulescu3}. The peculiarity of such equations with respect to the matrix difference equations is that in this case the unknowns are elements of a given abstract vector space $V$ whereas the \textquotedblleft coefficients \textquotedblright are linear operators acting on $V$.
We clarify this important point by introducing the following linear second order operator difference equation we are going to investigate and solve in this paper:

\begin{equation}\label{Cauchyprb}
      Y_{n+2}=\mathcal{L}_0(n)Y_n+\mathcal{L}_1(n)Y_{n+1}
      ,\qquad n\in \mathbb{N}^*.\end{equation}

This equation lives in a vector space $V$ over a field $\mathbb{F}$ and the nature will not be further specified.
 The null vector of $V$ is denoted by $\underline{0}$, while
 $\mathcal{L}_0(n)$  and $\mathcal{L}_1(n), n\in \mathbb{N}$ are two families of linear operators acting in $V$.
  To keep the investigation at a general level as much as possible, we do not introduce any constraint concerning the commutability of these operators in and between the two classes.

The aim of this paper is to build the general solution of equation (\ref{Cauchyprb}) when its coefficients are nonconstant and noncommutative. This means to give a noniterative expression of  $Y_n$, for a generic $n$, in terms of an appropriate ordered sequence of operators  $\mathcal{L}_0(n)$  and $\mathcal{L}_1(n)$. Of course this expression will contain two "summation constants" to be specified starting from given initial conditions.

The main result will be derived by mathematical induction in Section $3$ and some applications will be presented in the successive two sections.
%
%
\section{Mathematical preliminaries}

We  start by recalling that when $\forall n\in \mathbb{N}^* $ $\mathcal{L}_0(n)\equiv\mathcal{L}_0$ and $\mathcal{L}_1(n)\equiv\mathcal{L}_1$
 the notation
   $\{\mathcal{L}_0^{(u)}\mathcal{L}_1^{(v)}\}$ with $ u, v\in\mathbb{N}^{*}$  expresses the sum of all the  $ \left(\begin{array}{c}
                                u+v\\ m
                                \end{array} \right)$
 possible distinct permutations of $u$ $\mathcal{L}_0$-factors and  $v$ $\mathcal{L}_1$-factors, where  $m:=\min(u,v)$ \cite{Jivulescu1}.
 For example $\{\mathcal{L}_0^{(1)}\mathcal{L}_1^{(1)}\}=\mathcal{L}_0\mathcal{L}_1+\mathcal{L}_1\mathcal{L}_0$
 and we define $\{\mathcal{L}_0^{(0)}\mathcal{L}_1^{(0)}\}:=I$ putting consistently $ \left(\begin{array}{c}
                                0\\ 0
                                \end{array} \right)=1$.
It is possible to convince ourselves that any additive contribution to $\{\mathcal{L}_0^{(u)}\mathcal{L}_1^{(v)}\}$ may be represented as a formal product of $r (r\geq 1)$ powers of $\mathcal{L}_0$ alternated with $r$ powers of  $\mathcal{L}_1$, of the form
\begin{equation}\label{notatie}\mathcal{L}_0^{\tau_1}\mathcal{L}_1^{s_1}\mathcal{L}_0^{\tau_2}\mathcal{L}_1^{s_2}\ldots \mathcal{L}_0^{\tau_r}\mathcal{L}_1^{s_r}.
\end{equation}
The parameter $r$, hereforth called the length of the expression (\ref{notatie}), is an integer number running from $1$ to $m+1\equiv r_M$.
The $2r$ integers exponents
$\tau_1,\ldots, \tau_r$ and $s_1,\ldots, s_r$
are numbers fulfilling the following conditions
\begin{equation}\label{cond}\left\{\begin{array}{rl}
&\sum\limits_{i=1}^{r}\tau_i=u
\\&\sum\limits_{i=1}^{r}s_i=v
\\&s_1\ge 1, \tau_1\ge 0, \tau_r\ge 1, s_r\ge 0
\\&\tau_i\geq 1,s_i\geq 1, \quad i=2,\ldots, r-1
\end{array}\right. .\end{equation}
Hereafter the symbol $S^r_{u,v}$ represents the set of all the possible pairs of $r$-uples $\bar\tau=(\tau_1,\ldots,\tau_r)$ and $\bar s=(s_1,\ldots,s_r)$ satisfying eq. (\ref{cond}).
Expanding the symbol $\{\mathcal{L}_0^{(u)}\mathcal{L}_1^{(v)}\}$ in accordance to its definition
we get contributions possessing all the compatible lengths and for any length all the possible terms obtained in accordance with eq. (\ref{cond}).
It is possible to show that the number
of additive contributions of equal length $r$ to   $\{\mathcal{L}_0^{(u)}\mathcal{L}_1^{(v)}\}$ is given by $\left(\begin{array}{c}
                                u\\ r-1
                                \end{array} \right)\left(\begin{array}{c}
                                v\\ r-1
                                \end{array} \right)$ so that using the Vandermonde identity \cite{murray} \begin{equation}\sum\limits_{i=1}^{m+1}\left(\begin{array}{c}
                                u\\ i-1
                                \end{array} \right)\left(\begin{array}{c}
                                v\\ m+1-i
                                \end{array} \right)=\left(\begin{array}{c}
                                u+v\\ m
                                \end{array} \right)\end{equation}
 we derive  the following length-based representation of
$\{\mathcal{L}_0^{(u)}\mathcal{L}_1^{(v)}\}$
\begin{equation}\label{ordd}
\{\mathcal{L}_0^{(u)}\mathcal{L}_1^{(v)}\}=\sum\limits_{r=1}^{m+1}\sum\limits_{(\bar\tau,\bar s)\in S^r_{u,v}}\mathcal{L}_0^{\tau_1}\mathcal{L}_1^{s_1}\mathcal{L}_0^{\tau_2}\mathcal{L}_1^{s_2}\ldots \mathcal{L}_0^{\tau_r}\mathcal{L}_1^{s_r}
\end{equation}
where the summation
\begin{equation}
\sum\limits_{(\bar\tau,\bar s)\in S^r_{u,v}}\mathcal{L}_0^{\tau_1}\mathcal{L}_1^{s_1}\mathcal{L}_0^{\tau_2}\mathcal{L}_1^{s_2}\ldots \mathcal{L}_0^{\tau_r}\mathcal{L}_1^{s_r}\end{equation}
 runs over all the pairs $(\bar\tau,\bar s)$ compatible with the conditions (\ref{cond}).
The operator coefficients  appearing in eq. (\ref{Cauchyprb}),
depend on $n$ and do not commute with each other in general.  In such a case  powers like  $\mathcal{L}_0^{\tau_i}$ or $\mathcal{L}_1^{s_j}$  become ambiguous as soon as $\tau_i$ or $s_j$ exceeds 1.
Thus, to keep as much as possible the advantages of the notation adopted in expression (\ref{notatie}),
we introduce the following descending products of operators\\ \begin{equation}\label{prod}\coprod\limits_{p=p_1}^{p_r}X_p:=X(p_r)\ldots X(p_{j-1})\ldots X(p_1),\qquad p_1\in\mathbb{N}^{*}.\end{equation}
 postulating that
 $\coprod\limits_{p=p_1}^{p_r}X_p=I$ when $p_r<p_1$.
 For the sake of clarity, we stress that the symbol introduced in the left-hand side of eq. (\ref{prod}) singles out the inverted ordered product of the noncommuting $p_r$ operators $X_{p_1},\ldots ,X_{p_r}$.
 \par Exploiting eq. (\ref{prod}) we get rid of the ambiguity possessed by expression (\ref{notatie}) due to the noncommutativity of $\mathcal{L}_i, i\in\{0,1\}$ by substituting it with the following ordered operator $ [\mathcal{L}_0^{\tau_1}\mathcal{L}_1^{s_1}\mathcal{L}_0^{\tau_2}\mathcal{L}_1^{s_2}\ldots \mathcal{L}_0^{\tau_r}\mathcal{L}_1^{s_r}]_q$ defined as follows:
\begin{eqnarray}\label{formula}
&{\left[\mathcal{L}_0^{\tau_1}\mathcal{L}_1^{s_1}\mathcal{L}_0^{\tau_2}\mathcal{L}_1^{s_2}\ldots \mathcal{L}_0^{\tau_r}\mathcal{L}_1^{s_r}\right]}_q:=
\\&\nonumber \left[\coprod\limits_{i_1=0}^{\tau_1-1}\mathcal{L}_0\bigg(k_q-2i_1\bigg)\right]\left[\coprod\limits_{j_1=0}^{s_1-1}\mathcal{L}_1\bigg(k_q-2\tau_1-j_1\bigg)\right]\ldots\\\nonumber
& \left[\coprod\limits_{i_2=0}^{\tau_2-1}\mathcal{L}_0\bigg(k_q-2\tau_1-s_1-2i_2\bigg)\right]\left[\coprod\limits_{j_2=0}^{s_2-1}\mathcal{L}_1\bigg(k_q-2(\tau_1+\tau_2)-s_1-j_2\bigg)\right]
\ldots\\\nonumber&
\left[\coprod\limits_{i_r=0}^{\tau_r-1}\mathcal{L}_0\bigg(k_q-2\sum\limits_{l=1}^{\tau_{r-1}}\tau_l-
\sum\limits_{l=1}^{s_{r-1}}s_l-2i_r\bigg)\right]\left[\coprod\limits_{j_r=0}^{s_r-1}\mathcal{L}_1\bigg(k_q-2\sum\limits_{l=1}^{\tau_{r}}\tau_l-
\sum\limits_{l=1}^{s_{r-1}}s_l-j_r\bigg)\right].
\label{exp}
\end{eqnarray}
where \begin{equation}\label{k}k_q:=2u+v-q,\quad q=1,2.\end{equation}

In the next section we shall see that the peculiar  \textquotedblleft order-explosion\textquotedblright of eq. (\ref{notatie}) as given by eq. (\ref{formula}) provides a very useful
tool to write down the solution of a generic Cauchy problem associated to eq. (\ref{Cauchyprb}).
In the following
 the symbol $\{\mathcal{L}_0^{(u)}\mathcal{L}_1^{(v)}\}_q$ represents the sum (\ref{ordd}) where each additive contribution
 $\{\mathcal{L}_0^{(u)}\mathcal{L}_1^{(v)}\}$ is replaced
by the associated ordered operator that is
\begin{eqnarray}\label{ordperm} \{\mathcal{L}_0^{(u)}\mathcal{L}_1^{(v)}\}_q=\sum\limits_{r=1}^{m+1}\sum\limits_{(\bar\tau,\bar s)\in S^r_{u,v}}{\left[\mathcal{L}_0^{\tau_1}
\mathcal{L}_1^{s_1}\ldots \mathcal{L}_0^{\tau_r}\mathcal{L}_1^{s_r}\right]}_q.
\end{eqnarray}
To clarify the notation, we develop for  example $\{\mathcal{L}_0^2\mathcal{L}_1^1\}_1$:
\begin{eqnarray}\label{11} \{\mathcal{L}_0^{(2)}\mathcal{L}_1^{(1)}\}_1=&\sum\limits_{r=1}^2\sum\limits_{(\bar\tau,\bar s)\in S^r_{2,1}}
\left[ \mathcal{L}_0^{\tau_1}\mathcal{L}_1^{s_1}\mathcal{L}_0^{\tau_2}\mathcal{L}_1^{s_2}\ldots \mathcal{L}_0^{\tau_r}\mathcal{L}_1^{s_r}\right]_1
\\=&\sum\limits_{(\bar\tau,\bar s)\in S^1_{2,1}}[\mathcal{L}_0^{\tau_1}\mathcal{L}_1^{s_1}]_1+
\sum\limits_{(\bar\tau,\bar s)\in S^2_{2,1}}[\mathcal{L}_0^{\tau_1}\mathcal{L}_1^{s_1}\mathcal{L_0}^{\tau_2}\mathcal{L}_1^{s_2}]_1\nonumber
\\=&[\mathcal{L}_0^2\mathcal{L}_1]_1+ [\mathcal{L}_0^0\mathcal{L}_1^1\mathcal{L}_0^2\mathcal{L}_1^0]_1+[\mathcal{L}_0^1\mathcal{L}_1^1\mathcal{L}_0^1\mathcal{L}_1^0]_1.\nonumber
\end{eqnarray}
Since $k_1=4$, the three terms acquire the following explicit form:
%
\begin{description}
\item ${\left[\mathcal{L}_0^2\mathcal{L}_1^1\right]}_1=\left[\coprod\limits_{i_1=0}^{1}\mathcal{L}_0\bigg(4-2i_1\bigg)\right]\left[\coprod\limits_{j_1=0}^{0}\mathcal{L}_1\bigg(j_1\bigg)\right]=\mathcal{L}_0(4)\mathcal{L}_0(2)\mathcal{L}_1(0)$,
\item $[\mathcal{L}_0^0\mathcal{L}_1^1\mathcal{L}_0^2\mathcal{L}_1^0]_1=I\left[\coprod\limits_{j_1=0}^{0}\mathcal{L}_1\bigg(4-j_1\bigg)\right]\left[\coprod\limits_{i_2=0}^{1}\mathcal{L}_0\bigg(3-2i_2\bigg)\right]I=\mathcal{L}_1(4)\mathcal{L}_0(3)\mathcal{L}_0(1)$,

\item $[\mathcal{L}_0^1\mathcal{L}_1^1\mathcal{L}_0^1\mathcal{L}_1^0]_1=\mathcal{L}_0(4)\mathcal{L}_1(2)\mathcal{L}_0(1).$
\end{description}
We thus conclude that
\begin{eqnarray}\nonumber \{\mathcal{L}_0^2\mathcal{L}_1^1\}_1= \mathcal{L}_0(4)\mathcal{L}_0(2)\mathcal{L}_1(0)+\mathcal{L}_1(4)\mathcal{L}_0(3)\mathcal{L}_0(1)+\mathcal{L}_0(4)\mathcal{L}_1(2)\mathcal{L}_0(1)
.\end{eqnarray}


\section{The resolution of a Cauchy problem associated with eq. (\ref{Cauchyprb}) }
It is convenient to extend the definition of the operator $\{\mathcal{L}_0^{(u)}\mathcal{L}_1^{(v)}\}_q$ to negative integer values of $u$ and $v$, simply putting, when $u$ or $v$ are negative integers, $\{\mathcal{L}_0^{(u)}\mathcal{L}_1^{(v)}\}_qY=\underline{0}$, for any $Y$  which ${\mathcal{L}_0}$, ${\mathcal{L}_1}$ can legitimately act on. We are now ready to prove the following theorem constituting the main result of the paper:
\begin{theorem} The solution of the equation (\ref{Cauchyprb}) given that $Y_0=\underline{0}$ and $Y_1=B$ may be written down as
\begin{equation}\label{sol}Y_n=\sum\limits_{t=0}^{[\frac{|n-1|}{2}]}\{\mathcal{L}_0^t\mathcal{L}_1^{n-1-2t}\}_1B.\end{equation}
\end{theorem}
\textit{Proof}
\\
We prove (\ref{sol}) by mathematical induction.
For $ n=0$ and $n=1$, eq. (\ref{sol}) gives $Y_{0}=\underline{0}$ and $Y_1=B$ as expected. It is immediate to deduce $Y_2=B$ from both eq. (\ref{Cauchyprb}) and eq.(\ref{sol}).

Let's suppose now that the first $(n + 1)$ terms of the sequence solution of eq. (\ref{Cauchyprb}) are representable by
eq. (\ref{sol}). We must prove that whatever $n>0$ is
\begin{eqnarray}\label{concluzie}
&Y_{n+2}=\sum\limits_{t=0}^{[\frac{n+1}{2}]}\{\mathcal{L}_0^t\mathcal{L}_1^{n+1-2t}\}_1B
\\\nonumber&=\sum\limits_{t=0}^{[\frac{n+1}{2}]}\sum\limits_{r=1}^{m'+1}\sum\limits_{(\bar\tau,\bar s)\in S^r_{t,n+1-2t}}
[\mathcal{L}_0^{\tau_1}\mathcal{L}_1^{s_1}\ldots\mathcal{L}_0^{\tau_r}\mathcal{L}_1^{s_r}]_1B
\end{eqnarray}
where  $m'=\min\{t, n+1-2t\}$ satisfies eq. (\ref{Cauchyprb}). \\To this end we start by appropriately transforming
\begin{eqnarray}\label{l0}
&\mathcal{L}_0(n)Y_n=\mathcal{L}_0(n)\sum\limits_{t=0}^{[\frac{n-1}{2}]}\{\mathcal{L}_0^{t}\mathcal{L}_1^{n-1-2t}\}_1B
\\\nonumber&=\mathcal{L}_0(n)\sum\limits_{t=1}^{[\frac{n-1}{2}]+1}\{\mathcal{L}_0^{t-1}\mathcal{L}_1^{n+1-2t}\}_1B
\\\nonumber&=\mathcal{L}_0(n)\sum\limits_{t=1}^{[\frac{n+1}{2}]}\sum\limits_{r=1}^{m''+1}\sum\limits_{(\bar\tau,\bar s)\in S^r_{t-1,n+1-2t}}[\mathcal{L}_0^{\tau_1}\mathcal{L}_1^{s_1}\ldots\mathcal{L}_0^{\tau_r}\mathcal{L}_1^{s_r}]_1B
\end{eqnarray}
where $m''=\min\{t-1,n+1-2t\}\leq m'$ and we have used the fact that $[\frac{n-1}{2}]+1=[\frac{n+1}{2}]$.
In the following we are going to show that for every $r$, $1\leq r\leq m''+1$ the following operator relation holds:
\begin{eqnarray}\label{I}
&\mathcal{L}_0(n)\sum\limits_{(\bar\tau,\bar s)\in S^r_{t-1,n+1-2t}}
[\mathcal{L}_0^{\tau_1}\mathcal{L}_1^{s_1}\ldots\mathcal{L}_0^{\tau_r}\mathcal{L}_1^{s_r}]_1\\\nonumber&=
\sum\limits_{(\bar\tau,\bar s)\in S^r_{t,n+1-2t/\tau_1>0}}[\mathcal{L}_0^{\tau_1}\mathcal{L}_1^{s_1}\ldots\mathcal{L}_0^{\tau_r}\mathcal{L}_1^{s_r}]_1
\end{eqnarray}
where the right-hand side summation symbol means that the ordered operators $[\mathcal{L}_0^{0}\mathcal{L}_1^{s_1}\ldots\mathcal{L}_0^{\tau_r}\mathcal{L}_1^{s_r}]_1$ are not included.\\
We point out that $\mathcal{L}_0(n)[\mathcal{L}_0^{\tau_1}\mathcal{L}_1^{s_1}\ldots\mathcal{L}_0^{\tau_r}\mathcal{L}_1^{s_r}]_1$
generates permutations of the same length $r$ as $\mathcal{L}_0^{\tau_1}\mathcal{L}_1^{s_1}\ldots\mathcal{L}_0^{\tau_r}\mathcal{L}_1^{s_r}$ with $\tau_1$ at least $1$ necessarily. In addition we stress that the range $\{1,2,\ldots,m''+1\}$ of $r$ is compatible with the pair $u=t, v=n+1-2t$.
First, using eq. (\ref{formula}) and taking into account that in this case $k_1=n-2$, the left-hand side term of eq. (\ref{I}) may be developed as follows:
\begin{eqnarray}\label{gen}
&\mathcal{L}_0(n)
\sum\limits_{(\bar\tau,\bar s)\in S^r_{t-1,n+1-2t}}
\left[\coprod\limits_{i_1=0}^{\tau_1-1}\mathcal{L}_0\bigg(n-2-2i_1\bigg)\right]\left[\coprod\limits_{j_1=0}^{s_1-1}\mathcal{L}_1\bigg(n-2-2\tau_1-j_1\bigg)\right]
\\\nonumber&\times\left[\coprod\limits_{i_2=0}^{\tau_2-1}\mathcal{L}_0\bigg(n-2-2\tau_1-s_1-2i_2\bigg)\right]\left[\coprod\limits_{j_2=0}^{s_2-1}\mathcal{L}_1\bigg(n-2-2(\tau_1+\tau_2)-s_1-j_2\bigg)\right]\ldots\\
\nonumber&\times\left[\coprod\limits_{i_r=0}^{\tau_r-1}\mathcal{L}_0\bigg(n-2-2\sum\limits_{l=1}^{\tau_{r-1}}\tau_l-
\sum\limits_{l=1}^{s_{r-1}}s_l-2i_r\bigg)\right]\left[\coprod\limits_{j_r=0}^{s_r-1}\mathcal{L}_1\bigg(n-2-2\sum\limits_{l=1}^{\tau_{r}}\tau_l-
\sum\limits_{l=1}^{s_{r-1}}s_l-j_r\bigg)\right].
\end{eqnarray}
Shifting up the running index $i_1$ by $1$ that is putting $I_1=:i_1+1$, we transform the first $\mathcal{L}_0$-segment
$\left[\coprod\limits_{i_1=0}^{\tau_1-1}\mathcal{L}_0\bigg(n-2-2i_1\bigg) \right]$ appearing in eq. (\ref{gen}) into
$\coprod\limits_{I_1=1}^{\tau_1}\mathcal{L}_0\bigg(n-2I_1\bigg)$, which in turn may be rewritten as
$\coprod\limits_{I_1=0}^{\tau_1}\mathcal{L}_0\bigg(n-2I_1\bigg)$ by multiplication with $\mathcal{L}_0(n)$.


Introducing the positive index $T_1=:\tau_1+1$ the previous expression (\ref{gen}) may be cast in the following form
\begin{eqnarray}\label{19}
&\sum\limits_{(\bar\tau,\bar s)\in S^r_{t,n+1-2t/\tau_1=T_1}}
\left[\coprod\limits_{I_1=0}^{T_1-1}\mathcal{L}_0\bigg(n-2I_1\bigg)\right]\left[\coprod\limits_{j_1=0}^{s_1-1}\mathcal{L}_1\bigg(n-2T_1-j_1\bigg)\right]
\\\nonumber&\times\left[\coprod\limits_{i_2=0}^{\tau_2-1}\mathcal{L}_0\bigg(n-2T_1-s_1-2i_2\bigg)\right]\left[\coprod\limits_{j_2=0}^{s_2-1}\mathcal{L}_1\bigg(n-2(T_1+\tau_2)-s_1-j_2\bigg)\right]
\ldots\\\nonumber
\\\nonumber&\times\left[\coprod\limits_{i_r=0}^{\tau_r-1}\mathcal{L}_0\bigg(n-2T_1-2\sum\limits_{l=2}^{\tau_{r-1}}\tau_l-\sum\limits_{r=1}^{s_{q-1}}s_r-2i_r\bigg)
\right]\left[\coprod\limits_{j_r=0}^{s_r-1}\mathcal{L}_1\bigg(n-2T_1-2\sum\limits_{l=2}^{\tau_{r}}\tau_l-\sum\limits_{l=1}^{s_{r-1}}s_l-j_r\bigg)\right].
\end{eqnarray}
where each ordered contribution still has length $r$ and the passage from $t-1$ to $t$ arises from the \textquotedblleft absorbtion\textquotedblright of $\mathcal{L}_0(n)$.
Eq. (\ref{19}) is exactly the development of the right-hand side term of eq. (\ref{I}), where $T_1$ plays the role of $\tau_1$ and then eq. (\ref{l0}) may be given the following aspect:
\begin{eqnarray}\label{l00}
\mathcal{L}_0(n)Y_{n}=
&\sum\limits_{t=1}^{[\frac{n+1}{2}]}
\sum\limits_{r=1}^{m''+1}
\sum\limits_{(\bar\tau,\bar s)\in S^r_{t,n+1-2t/\tau_1>0}}
[\mathcal{L}_0^{\tau_1}\mathcal{L}_1^{s_1}\ldots\mathcal{L}_0^{\tau_r}\mathcal{L}_1^{s_r}]_1B
\end{eqnarray}
To establish a connection between eq. (\ref{l00}) and eq. (\ref{concluzie}) we extract from the latter expression the ordered operators beginning with $\mathcal{L}_0^{\tau_1}$, with $\tau_1>0$
\begin{equation}\label{a}
\sum\limits_{t=1}^{[\frac{n+1}{2}]}\sum\limits_{r=1}^{m'+1}
\sum\limits_{(\bar\tau,\bar s)\in S^r_{t,n+1-2t/\tau_1>0}}
[\mathcal{L}_0^{\tau_1}\mathcal{L}_1^{s_1}\ldots\mathcal{L}_0^{\tau_r}\mathcal{L}_1^{s_r}]_1.
\end{equation}
Here $t=0$ is excluded  since it implies $m'=0$ and then necessarily $\tau_1=0$.
If $m'=\min(t, n+1-2t)=t$ the correspondent highest
length $r_M=t+1$ is indeed certainly incompatible with the condition $\tau_1>0$.
 Thus we are legitimated to substitute $m'$ with $m''=\min(t-1,n+1-2t)$ in the expression (\ref{a}). These arguments prove that $\mathcal{L}_0(n)Y_n$ captures all the ordered operators appearing in eq. (\ref{concluzie}) effectively beginning with an $\mathcal{L}_0$-segment ($\tau_1>0$). \\In the following in view of eq. (\ref{Cauchyprb}) we are going to prove that $\mathcal{L}_1(n)Y_{n+1}$ coincides with the sum of all the other contributions in eq. (\ref{concluzie}) namely all those ordered operators beginning with $\mathcal{L}_0^{\tau_1}(n)$ with $ \tau_1=0$.
To this end we exploit the inductive hypothesis writing down the action of $\mathcal{L}_1(n)$ on $Y_{n+1}$ as follows:
\begin{eqnarray}\label{l1}
&\mathcal{L}_1(n)Y_{n+1}=\mathcal{L}_1(n)\sum\limits_{t=0}^{[\frac{n}{2}]}\{\mathcal{L}_0^t\mathcal{L}_1^{n-2t}\}_1B
\\&=\mathcal{L}_1(n)\sum\limits_{t=0}^{[\frac{n}{2}]}\sum\limits_{r=1}^{m'''+1}
\sum\limits_{(\bar\tau,\bar s)\in S^r_{t,n-2t}}
[\mathcal{L}_0^{\tau_1}\mathcal{L}_1^{s_1}\ldots\mathcal{L}_0^{\tau_r}\mathcal{L}_1^{s_r}]_1B.\nonumber
\end{eqnarray}
where $m'''=\min\{t,n-2t\}$.

We firstly consider the terms  of increasing length $r$ in $Y_{n+1}$, having $\tau_1=0$. Taking into account that  $k_1=n-1$ when $r=1$  we get  from  eq. (\ref{formula}) the unique contribution ($t=0$)
\begin{eqnarray}\label{primul}
&\mathcal{L}_1(n)
\sum\limits_{(\bar\tau,\bar s)\in S^1_{0,n/\tau_1=0}}
[\mathcal{L}_0^{\tau_1}\mathcal{L}_1^{s_1}]_1= \mathcal{L}_1(n)[\mathcal{L}_0^{0}\mathcal{L}_1^{n}]_1\\\nonumber&=
\mathcal{L}_1(n)\left[\coprod\limits_{j_1=0}^{n-1}\mathcal{L}_1\bigg(n-1-j_1\bigg)\right]=\mathcal{L}_1(n)\left[\coprod\limits_{J_1=1}^n\mathcal{L}_1\bigg(n-J_1\bigg)\right]\\\nonumber&=\left[\coprod
\limits_{J_1=0}^n\mathcal{L}_1\bigg(n-J_1\bigg)\right]=
\{\mathcal{L}_0^{0}\mathcal{L}_1^{n+1}\}_1
\end{eqnarray}
where we have replaced $j_1$ by $J_1-1$. This operator applied to $B$ is present in eq. (\ref{concluzie}), $(t=0)$.
\\Now we are going to prove that a generic term of $Y_{n+1}$ having length $r>1$ and still with $\tau_1=0$ satisfies the property
\begin{eqnarray}\label{III}
&
\mathcal{L}_1(n)
\sum\limits_{(\bar\tau,\bar s)\in S^r_{t,n-2t/\tau_1=0}}
[\mathcal{L}_0^{\tau_1}\mathcal{L}_1^{s_1}\ldots\mathcal{L}_0^{\tau_r}\mathcal{L}_1^{s_r}]_1=
\\\nonumber&
\sum\limits_{(\bar\tau,\bar s)\in S^r_{t,n+1-2t/\tau_1=0,s_1>1}}
[\mathcal{L}_0^{\tau_1}\mathcal{L}_1^{s_1}\ldots\mathcal{L}_0^{\tau_r}\mathcal{L}_1^{s_r}]_1.
\end{eqnarray}
Expanding the left-hand side of eq. (\ref{III}) we indeed get
\begin{eqnarray}
&\mathcal{L}_1(n)
\sum\limits_{(\bar\tau,\bar s)\in S^r_{t,n-2t/\tau_1=0}}
 \left[\coprod\limits_{i_1=0}^{\tau_1-1}\mathcal{L}_0\bigg(n-1-2i_1\bigg)\right]\left[\coprod\limits_{j_1=0}^{s_1-1}\mathcal{L}_1\bigg(n-1-2\tau_1-j_1\bigg)\right]
\\\nonumber&\times\left[\coprod\limits_{i_2=0}^{\tau_2-1}\mathcal{L}_0\bigg(n-1-2\tau_1-s_1-2i_2\bigg)\right]\left[\coprod\limits_{j_2=0}^{s_2-1}\mathcal{L}_1\bigg(n-1-2(\tau_1+\tau_2)
-s_1-j_2\bigg)\right]\ldots\\\nonumber
\\\nonumber&\times\left[\coprod\limits_{i_r=0}^{\tau_r-1}\mathcal{L}_0\bigg(n-1-2\sum\limits_{l=1}^{\tau_{r-1}}\tau_l-\sum\limits_{l=1}^{s_{r-1}}s_l-2i_r\bigg)\right]
\left[\coprod\limits_{j_r=0}^{s_r-1}\mathcal{L}_1\bigg(n-1-2\sum\limits_{l=1}^{\tau_{r}}\tau_l-\sum\limits_{l=1}^{s_{r-1}}s_l-j_r\bigg)\right]
\end{eqnarray}
which, putting   $j_1+1=J_1, s_1+1=S_1$, yields
\begin{eqnarray}
\label{26}
&\sum\limits_{(\bar\tau,\bar s)\in S^r_{t,n-2t/\tau_1=0,s_1=S_1>1}}
 \left[\coprod\limits_{i_1=0}^{\tau_1-1}\mathcal{L}_0\bigg(n-2i_1\bigg)\right]\left[\coprod\limits_{J_1=0}^{S_1-1}\mathcal{L}_1\bigg(n-2\tau_1-J_1\bigg)\right]
\\\nonumber&\times\left[\coprod\limits_{i_2=0}^{\tau_2-1}\mathcal{L}_0\bigg(n-2\tau_1-S_1-2i_2\bigg)\right]\left[\coprod\limits_{j_2=0}^{s_2-1}\mathcal{L}_1\bigg(n-2(\tau_1+\tau_2)-S_1-j_2\bigg)\right]\ldots\\\nonumber
\\\nonumber&\left[\coprod\limits_{i_m=0}^{\tau_m-1}\mathcal{L}_0\bigg(n-2\sum\limits_{l=1}^{\tau_{r-1}}\tau_l-S_1-\sum\limits_{l=2}^{s_{r-1}}s_l-2i_r\bigg)\right]
\left[\coprod\limits_{j_r=0}^{s_r-1}\mathcal{L}_1\bigg(n-2\sum\limits_{l=1}^{\tau_{r}}\tau_l-S_1-\sum\limits_{q=2}^{s_{r-1}}s_l-j_r\bigg)\right].
\end{eqnarray}
Eq. (\ref{26}) is the expansion of the right-hand side of eq. (\ref{III}) where the increase from $(n-2t)$ to $(n+1-2t)$ is due to the \textquotedblleft absorbtion\textquotedblright of $\mathcal{L}_1(n)$ as given by eq. (\ref{26}) and
 the condition $s_1>1$  directly stems from the fact that by construction $r>1$ in eq. (\ref{III}).
Summing both members eq. (\ref{III}) in view of eq. (\ref{l1}) over $r>1$ and $t>1$ ($t=0\Rightarrow r=1$) yields
\begin{eqnarray}\nonumber
&\mathcal{L}_1(n)\sum\limits_{t=1}^{[\frac{n}{2}]}\sum\limits_{r=2}^{m'''+1}
    \sum\limits_{(\bar\tau,\bar s)\in S^r_{t,n-2t/\tau_1=0}}
[\mathcal{L}_0^{\tau_1}\mathcal{L}_1^{s_1}\ldots\mathcal{L}_0^{\tau_r}\mathcal{L}_1^{s_r}]_1=
\\\label{b}&    \sum\limits_{t=1}^{[\frac{n}{2}]}\sum\limits_{r=2}^{m'''+1}
    \sum\limits_{(\bar\tau,\bar s)\in S^r_{t,n+1-2t/\tau_1=0,s_1>1}}
[\mathcal{L}_0^{\tau_1}\mathcal{L}_1^{s_1}\ldots\mathcal{L}_0^{\tau_r}\mathcal{L}_1^{s_r}]_1
\end{eqnarray}
Note that since $m'''<m'$ each ordered operator of length $r>1$ compatible with $\{u=t,v=n-2t\}$ generates an ordered operator compatible with $\{u=t,v=n+1-2t\}$
increasing $s_1$ by $1$ thus rendering in the latter case $s_1>1$. This circumstance means that all the ordered operators appearing in  eq. (\ref{b})
determine only the set of all ordered operators having $r>1$ and compatible with the maximum length $m'+1$
under the constraints $\tau_1=0$ and $s_1>1$. Thus putting together eq. (\ref{primul}) and eq. (\ref{l1})  we claim to have proved that
\begin{eqnarray}\label{bb}
   &\mathcal{L}_1{(n)} \sum\limits_{t=0}^{[\frac{n}{2}]}\sum\limits_{r=1}^{m'''+1}
   \sum\limits_{(\bar\tau,\bar s)\in S^r_{t,n-2t/\tau_1=0}}
[\mathcal{L}_0^{\tau_1}\mathcal{L}_1^{s_1}\ldots\mathcal{L}_0^{\tau_r}\mathcal{L}_1^{s_r}]_1=\\\nonumber & \sum\limits_{t=0}^{[\frac{n+1}{2}]}\sum\limits_{r=1}^{m'+1}
\sum\limits_{(\bar\tau,\bar s)\in S^r_{t,n+1-2t/\tau_1=0,s_1>1}}
[\mathcal{L}_0^{\tau_1}\mathcal{L}_1^{s_1}\ldots\mathcal{L}_0^{\tau_r}\mathcal{L}_1^{s_r}]_1.
\end{eqnarray}
where $[\frac{n}{2}]$ may be substituted with $[\frac{n+1}{2}]$ exploiting the fact that for any $n$ odd the highest value of $t$ gives rise to contribution for which $\tau_1>0$.
We concentrate now on the action of $\mathcal{L}_1(n)$ on the generic terms of $Y_{n+1}$ in eq. (\ref{l1}) for which $\tau_1> 0$, that is
\begin{eqnarray}\label{27}
&\mathcal{L}_1(n)
\sum\limits_{(\bar\tau,\bar s)\in S^r_{t,n-2t/\tau_1>0}}
[\mathcal{L}_0^{\tau_1}\mathcal{L}_1^{s_1}\ldots \mathcal{L}_0^{\tau_{r}}\mathcal{L}_1^{s_{r}}]_1
 \end{eqnarray}
To this end
we observe that under the constraint $\tau_1>0$ the ordered operators as  given by eq. (\ref{l1})
in the length-based expansion of $Y_{n+1}$
do not contribute to the first member of eq. (\ref{27}) for $r_M=t+1, (m'''=t)$ in accordance with eq. (\ref{sol}). When instead $m'''<t$ then $m'''+2=(n+1-2t)+1=m'+1$
since in such a condition $m'=\min(t,n+1-2t)=n+1-2t$.
Considering that when $\tau_0=0$ and $s_0=1$ the operator $\mathcal{L}_1(n)$ may be represented as
\begin{equation}\label{l}
  \mathcal{L}_1(n)=\mathcal{L}^0_0(n)\mathcal{L}^1_1(n)=\left[\coprod\limits_{i_0=0}^{\tau_0-1}\mathcal{L}_0\bigg(n-2i_0\bigg)\right]\left[\coprod\limits_{j_0=0}^{s_0-1} \mathcal{L}_1\bigg(n-2\tau_0-j_0\bigg)\right]
\end{equation}
may be rewritten as follows:
\begin{eqnarray}\label{c}&
    \mathcal{L}_1(n)
    \sum\limits_{(\bar\tau,\bar s)\in S^r_{t,n-2t/\tau_1>0}}
   [\mathcal{L}_0^{\tau_1}\mathcal{L}_1^{s_1}\ldots \mathcal{L}_0^{\tau_{r}}\mathcal{L}_1^{s_{r}}]_1=\\\nonumber&
   \sum\limits_{(\bar\tau,\bar s)\in S^r_{t,n-2t/\tau_1>0}}
  [\mathcal{L}_0^{0}\mathcal{L}_1^{1}\ldots \mathcal{L}_0^{\tau_{r}}\mathcal{L}_1^{s_{r}}]_1
\end{eqnarray}
where $r$ runs from $1$ to $m'''+1$ in accordance with eq. (\ref{l1}).
We are interested in summing on $r$ both members of eq. (\ref{c}) getting:
\begin{eqnarray}\label{d}
&\sum\limits_{r=1}^{m'''+1}
\sum\limits_{(\bar\tau,\bar s)\in S^r_{t,n-2t/\tau_1>0}}
[\mathcal{L}_0^{0}\mathcal{L}_1^{1}\mathcal{L}_0^{\tau_1}\mathcal{L}_1^{s_1}\ldots\mathcal{L}_0^{\tau_r}\mathcal{L}_1^{s_r}]_1= \\\nonumber
  &\sum\limits_{r=2}^{m'''+2}
  \sum\limits_{(\bar\tau,\bar s)\in S^r_{t,n+1-2t/\tau_1=0,s_1=1}}
[\mathcal{L}_0^{\tau_1}\mathcal{L}_1^{s_1}\ldots\mathcal{L}_0^{\tau_r}\mathcal{L}_1^{s_r}]_1.
\end{eqnarray}
As already discussed just before eq. (\ref{l})
if $m'''=t$ then $r$ may be stopped at $m'''+1=m'+1$ whereas if $m'''<t$ then $m'''+2=m'+1$ as well.
Thus we are legitimated to write down that
\begin{eqnarray}\label{e}
   & \sum\limits_{t=0}^{[\frac{n}{2}]}\sum\limits_{r=1}^{m'''+1}\mathcal{L}_1{(n)}
   \sum\limits_{(\bar\tau,\bar s)\in S^r_{t,n-2t/\tau_1>0}}
[\mathcal{L}_0^{\tau_1}\mathcal{L}_1^{s_1}\ldots\mathcal{L}_0^{\tau_r}\mathcal{L}_1^{s_r}]_1=\\\nonumber &
\sum\limits_{t=1}^{[\frac{n+1}{2}]}\sum\limits_{r=2}^{m'+1}
\sum\limits_{(\bar\tau,\bar s)\in S^r_{t,n+1-2t/\tau_1=0,s_1=1}}
[\mathcal{L}_0^{\tau_1}\mathcal{L}_1^{s_1}\ldots\mathcal{L}_0^{\tau_r}\mathcal{L}_1^{s_r}]_1.
\end{eqnarray}
We have thus completed the demonstration of theorem 1 since we have shown that $\mathcal{L}_0(n)Y_{n}$ $(\mathcal{L}_1(n)Y_{n+1})$ when $Y_n$ $(Y_{n+1})$ is expressed by eq. (\ref{sol}) generates all and only the ordered operators of $Y_{n+2}$ as given by the same equation beginning with $\mathcal{L}_0^{\tau_1}(n)$ with  $\tau_1>0$ $ (\tau_1=0)$.

\begin{remark} Exploiting a treatment analogous to the one used to demonstrate Theorem 1, it is possible to prove that the solution of eq. (\ref{Cauchyprb}) given that
 $Y_0=A$ and $Y_1=0$
 may be written down, for any $n\geq 2$, as follows
\begin{equation}\label{sol2}
Y_n=\left[\sum\limits_{t=0}^{[\frac{n-2}{2}]}\{\mathcal{L}_0^{t+1}\mathcal{L}_1^{n-2-2t}\}_{2/\tau_r=1,s_r=0}\right]A
\end{equation}
where all the ordered contributions to be considered in the summation are those finishing with $\mathcal{L}_0^{1}\mathcal{L}_1^{0}$.
\end{remark}
To clarify the notation, we give the expression of $Y_5:$
\begin{eqnarray} Y_5=\left[\{\mathcal{L}_0^{1}\mathcal{L}_1^{3}\}_{2/\tau_r=1,s_r=0}+\{\mathcal{L}_0^{2}\mathcal{L}_1^{1}\}_{2/\tau_r=1,s_r=0}\right]A.
\nonumber\end{eqnarray}
The constraints extract from the first term only one ordered operator of length $2$ and from the second one, two ordered contributions
only, of the same length $2$:
\begin{eqnarray}Y_5=\left[\mathcal{L}_1{(3)}\mathcal{L}_1{(2)}\mathcal{L}_1(1)\mathcal{L}_0(0)+\mathcal{L}_0(3)\mathcal{L}_1(1)\mathcal{L}_0(0)+\mathcal{L}_1(3)\mathcal{L}_0(2)\mathcal{L}_0(0)\right]A.\nonumber\end{eqnarray} Since the two Cauchy problems we solved give rise to two independent solutions of eq. (\ref{Cauchyprb}), its general solution, for any $n\geq 2$, may be written down exploiting eqs. (\ref{sol}) and  (\ref{sol2})
\begin{eqnarray} Y_n&=\left[\sum\limits_{t=0}^{[\frac{n-2}{2}]}\{\mathcal{L}_0^{t+1}\mathcal{L}_1^{n-2-2t}\}_{2/\tau_r=1,s_r=0}\right]A+\sum\limits_{t=0}^{[\frac{|n-1|}{2}]}\{\mathcal{L}_0^t\mathcal{L}_1^{n-1-2t}\}_1B.
\end{eqnarray}

\section{Applications}
In this section we show the effectiveness of eqs. (\ref{sol}) and (\ref{sol2}) by exactly solving a non-trivial example of eq. (\ref{Cauchyprb})
formulated in a complex vectorial space $V$ of even dimension $N$. Following Dirac denote by $|v\rangle$ a generic vector of $V$ and by $\langle v|w\rangle:=\langle w|v\rangle^*$ the scalar product between the two vectors $|v\rangle$ and $|w\rangle$ belonging to $V$.
If an orthonormal basis $\mathcal{B}$ of $V$ is composed of the $N$ vectors: $|1\rangle$ ,$|2\rangle , \ldots, |i\rangle, \ldots, |j\rangle, \ldots, |N\rangle$ such that $\langle i|j\rangle=\delta_{ij}$, let's introduce the transition operator $T_{ij}:=|i\rangle\langle j|$ acting upon a generic vector $|i'\rangle$ of $B$ as  follows: $T_{ij}|i'\rangle=\delta_{ji'}|i\rangle$.  We define the following linear operators acting on $V$ where the $N/2$ coefficients $\{c_i, i=1,2,\ldots, N/2\}$ are complex numbers:
 \begin{subequations}
\begin{equation}\label{m+} M_{+}=\sum\limits_{i=1}^{N/2}c_iT_{i,N-i+1}=\sum\limits_{i=1}^{N/2}c_i|i\rangle\langle N-i+1|
 \end{equation}
\begin{equation}\label{m-} M_{-}=(M_{+})^{\dagger}=\sum\limits_{i=1}^{N/2}c_i^*|N-i+1\rangle\langle i|
 \end{equation}
  \end{subequations}
   \begin{subequations}
\begin{equation}\label{d+} D_{+}=\sum\limits_{i=1}^{N/2}|c_i|^2|i\rangle\langle i|
 \end{equation}
\begin{equation}\label{d-} D_{-}=\sum\limits_{i=N/2+1}^{N}|c_{N-i+1}|^2|i\rangle\langle i|
 \end{equation}
  \end{subequations}
 \begin{subequations}\begin{equation}
\label{m} M_0=D_{+}-D_{-}
\end{equation}\begin{equation}\label{dn}
 D_{N}=D_{+}+D_{-}.
 \end{equation}\end{subequations}
 When $|c_i|=\rho_i=\rho$  whatever $i$ is it is easy to check the following properties:
  \begin{subequations}
 \begin{equation}\label{46}M_{\pm}^2|v\rangle=\underline{0},\qquad \forall |v\rangle\in V
 \end{equation}
 \begin{eqnarray}\label{47} [M_{+},M_{-}]=M_0
 \\\label{48}M_{\pm}M_{\mp}=D_{\pm}
 \end{eqnarray}
  \end{subequations}
  \begin{subequations}
 \begin{equation}\label{49}M_{-}D_{+}=\rho^2M_{-}\Rightarrow D_{+}M_{+}=\rho^2M_{+}
 \end{equation}
  \begin{equation}\label{410}M_{+}D_{+}|v\rangle=\underline{0}\Rightarrow D_{+}M_{-}|v\rangle=\underline{0},\qquad \forall |v\rangle\in V
 \end{equation}
 \begin{equation}\label{411}M_{-}D_{-}|v\rangle=\underline{0}\Rightarrow D_{-}M_{+}|v\rangle=\underline{0},\qquad \forall |v\rangle\in V
 \end{equation}
 \begin{equation}\label{412}M_{+}D_{-}=\rho^2M_{+}\Rightarrow D_{-}M_{-}=\rho^2M_{-}.
 \end{equation}
  \end{subequations}

We point out that eqs. (\ref{46})-(\ref{48}) and (\ref{410})-(\ref{411}) hold without any restriction on the coefficients $c_n$
and that when $N=2$ and the corresponding unique coefficient $c_1=1$  $M_{+}, M_{-}$ and $M_0$ may be traced back to the well-known Pauli matrices. In the following analysis, we will make use only of the properties (\ref{46})-(\ref{412}), rather than using the explicit representation given by eqs. (\ref{m+})-(\ref{dn}). It is therefore worthy to underline that generally speaking, other sets of $M_{-}$ and $D_{-}$ matrices fulfilling the properties (\ref{m+})-(\ref{412}) exist.\\ The scope of this section is to exploit eqs. (\ref{sol}) and (\ref{sol2}) to provide the closed form of the following general Cauchy problem
\begin{subequations}
 \begin{equation}\label{Cauchyprbgen}
      Y_{n+2}=\mathcal{L}_0(n)Y_n+\mathcal{L}_1(n)Y_{n+1},\qquad n\in \mathbb{N}^*,
     \end{equation}
     \begin{equation}
      Y_0=\overline{Y}_0,\qquad Y_1=\overline{Y}_1, \end{equation}
      \end{subequations}
where $Y_n$ is a vector of $V$, $\overline{Y}_0,\overline{Y}_1$ are initial conditions fixed at will in $V$ and the generally noncommuting coefficients operators  $\mathcal{L}_0(n)$ and $\mathcal{L}_1(n), n\in \mathbb{N}^{*}$ are defined as functions of $n$, as follows:
 \begin{equation}\label{414}\mathcal{L}_0(n)=\left\{\begin{array}{rl}
     & M_{+}  \quad n\quad \mathrm{even},\\
      &M_{-}  \quad n\quad \mathrm{odd},\end{array}\right.\end{equation}

\begin{equation}\label{415}\mathcal{L}_1(n)=\left\{\begin{array}{rl}
      &M_{-}  \quad n\quad \mathrm{even},\\
      &M_{+} \quad n\quad \mathrm{odd}.\end{array}\right.\end{equation}
To this end it is enough to consider the two initial conditions
\begin{equation}\label{II}
   (\mathrm{I}) \quad Y_0=\underline{0},\quad Y_1=\overline{Y}_1.
\end{equation}
\begin{equation}\label{IIII}
   (\mathrm{II})\quad Y_0=\overline{Y}_0, \quad Y_1=\underline{0}
\end{equation}
\subsection{Solving the Cauchy problem (I)}
 Consider the operator $[ \mathcal{L}_0^{\tau_1}\mathcal{L}_1^{s_1}\mathcal{L}_0^{\tau_2}\mathcal{L}_1^{s_2}\ldots \mathcal{L}_0^{\tau_r}\mathcal{L}_1^{s_r}]_1$ and observe that as consequence of eqs. (\ref{46}), (\ref{414}), (\ref{415})
it vanishes as soon as one among the $r$ integer exponents $\tau_i$ is greater than 1.
 In this case eq. (\ref{sol}) indeed
 exhibits \textquotedblleft $\mathcal{L}_0$-segments\textquotedblright  containing $M_{\pm}^2$. Moreover if $r>1$ then $s_1\tau_1>1$ and  the operator $[\mathcal{L}_0^{\tau_1}\mathcal{L}_1^{s_1}\mathcal{L}_0^{\tau_2}\mathcal{L}_1^{s_2}\ldots \mathcal{L}_0^{\tau_r}\mathcal{L}_1^{s_r}]_1$
 vanishes because it includes
 the product
 \begin{equation}\ldots \mathcal{L}_1(k-2\tau_1-s_1+1)\mathcal{L}_0(k-2\tau_1-s_1)\ldots
 \end{equation}
 which coincides with $M_{\pm}^2$ in view of eqs. (\ref{414}) and (\ref{415}).
 Summing up, when the length $r$ of the operator $[\mathcal{L}_0^{\tau_1}\mathcal{L}_1^{s_1}\mathcal{L}_0^{\tau_2}\mathcal{L}_1^{s_2}\ldots \mathcal{L}_0^{\tau_r}\mathcal{L}_1^{s_r}]_1$
 exceeds $1$, its ordered expression given by eq. (\ref{formula}) vanishes.
  This property, of course, strictly related to the operators $\mathcal{L}_0,\mathcal{L}_1$
 greatly simplifies eq. (\ref{sol}) where the ordered values of $t$ of interest become $t=0$ (for any $n$) and $t=1$ (for $n\geq 4$).
 The reason is that $t>1$ and $r=1$ requires $\tau_1>1$ and then leads to a vanishing operator, as previously discussed.

We thus concentrate ourselves on the evaluation of the expression of two operators $\{\mathcal{L}_0^0\mathcal{L}_1^{n-1}\}_1$ and $\{\mathcal{L}_0^1\mathcal{L}_1^{n-3}\}_1$.
 For the first one we have $r=1$ and
 \begin{equation}\label{418}\{\mathcal{L}_0^0\mathcal{L}_1^{n-1}\}_1= [\mathcal{L}_0^0\mathcal{L}_1^{n-1}]_1=\mathcal{L}_1(n-2)\ldots \mathcal{L}_1(0)=M_{*}\ldots M_{+}M_{-}\end{equation}
where $*$ is $-(+)$ if $n$ is even (odd).  The second and third equalities stem from the application of eq. (\ref{formula})
and eq. (\ref{415}) respectively.
Taking into account eq. (\ref{48}) and the constraints on the complex entries $c_i$, we finally get
\begin{equation}\label{419}\{\mathcal{L}_0^0\mathcal{L}_1^{n-1}\}_1=\left\{\begin{array}{rl}
     \rho^{n-3} D_{+}&  \quad n\quad \mathrm{odd},\\
      \rho^{n-2}M_{-}&  \quad n\quad \mathrm{even}.\end{array}\right.\end{equation}
When $t=1 (n\geq 4),$ the maximal length $r_M$ compatible with $t$ is $ r_M=2$ and
\begin{equation}\label{420}
   \{\mathcal{L}_0^1\mathcal{L}_1^{n-3}\}_1=[\mathcal{L}_0^1\mathcal{L}_1^{n-3}]_1.
\end{equation}
Then it is easy to check that
\begin{equation}\label{421}[\mathcal{L}_0^1\mathcal{L}_1^{n-3}]_1=\mathcal{L}_0(n-2)\mathcal{L}_1(n-4)\ldots\mathcal{L}_1(0) =\left\{\begin{array}{rl}
     &\rho^{n-4} D_{+} \quad n\quad \mathrm{even},\\
      &\rho^{n-3}M_{-}  \quad n\quad \mathrm{odd}.\end{array}\right.\end{equation}
 Substituting eqs. (\ref{419}) and (\ref{420})
into eq. (\ref{sol}), taking into account that all the contributions from any $t>1$ vanish, the solution of the Cauchy problem (\ref{I})
may be cast in the following closed and explicit form
\begin{equation}\label{422}Y_n^{(I)}=
    \left\{\begin{array}{rl}
     \underline{0} & \quad n=0,\\
    \overline{Y}_1&\quad n=1,\\
     M_{-}\overline{Y}_1&\quad n=2,\\
    (\rho^{n-4}D_{+}+{\rho}^{n-2}M_{-})\overline{Y}_1&\quad n>2, \mathrm{even},\\
      \rho^{n-3(}D_{+}+M_{-})\overline{Y}_1 &\quad n>2, \mathrm{odd}.
      \end{array}\right.
\end{equation}
In passing we note that, when $\rho_i=\rho=1$ for any $i$, $Y_n=(D_{+}+M_{-})\overline{Y}_1$, whatever $n>2$ is which means that we obtain a very simple constant solution in this case.
\subsection{Solving the Cauchy problem (II)}
 We make use of eq. (\ref{sol2}) observing that since for $n>2$
\begin{equation}\label{423}
    \{\mathcal{L}_0^1\mathcal{L}_1^{n-2}\}_{2/\tau_r=1,s_r=0}=[\mathcal{L}_1^{n-2}\mathcal{L}_0^1=\mathcal{L}_1(n-2)\ldots\mathcal{L}_1(1)\mathcal{L}_0(0)]_2=0
\end{equation}
In addition we have
for any $n\geq 4$
\begin{equation}\label{424}
   \{\mathcal{L}_0^2\mathcal{L}_1^{n-4}\}_{2/\tau_r=1,s_r=0}=[\mathcal{L}_0^1\mathcal{L}_1^{n-4}\mathcal{L}_0^1]_2=\mathcal{L}_0(n-2)\mathcal{L}_1(n-4)\ldots\mathcal{L}_1(1)\mathcal{L}_0(0)=0
\end{equation}
Thus the solution of the Cauchy problem (II) under scrutiny
assumes the following form
\begin{equation}\label{4255}Y_n^{(II)}=
    \left\{\begin{array}{rl}
     \overline{Y}_0 &\quad n=0,\\
     \underline{0}& \quad n=1,\\
    \mathcal{L}_0\overline{Y}_0 & \quad n=2,\\
  \underline{0} & \quad n>2.
      \end{array}\right.
\end{equation}
By direct substitution it is easy to check that this very simple sequence of vectors of $V$ is the (unique) solution of the Cauchy problem (II).  Exploiting eqs. (\ref{422}) and (\ref{4255}), we write down the solution $Y_n$ of the general Cauchy problem as follows:
\begin{equation}\label{425}Y_n=
    \left\{\begin{array}{rl}
     \overline{Y}_0& \quad n=0,\\
     \overline{Y}_1&\quad n=1,\\
    \mathcal{L}_0(0)\overline{Y}_0 +M_{-}\overline{Y}_1 &\quad n=2,\\
  (\rho^{n-4}D_{+}+\rho^{n-2}M_{-})\overline{Y}_1 & \quad n>2, \mathrm{even},\\
  \rho^{n-3}(D_{+}+M_{-})\overline{Y}_1  & \quad n>2, \mathrm{odd}.
      \end{array}\right.
\end{equation}
Of course, interpreting the initial conditions $\overline{Y}_0$ and $\overline{Y}_1$
as playing the role of \textquotedblleft summation constants\textquotedblright, eq. (\ref{425}) may be also refereed to as the general solution of difference equation (\ref{Cauchyprbgen}).
\\In conclusion we stress that the target of this section that is to find the general solution
of a selected difference equation exploiting eqs. (\ref{sol}) (\ref{sol2}) and has been reached by eq. (\ref{425}).
We do believe that solving this not trivial toy-difference equation represents a good test to appreciate the potential and practical value of the theory we have reported in the previous sections of the paper.

\section{Concluding remarks}
The applicative potentialities of eq. (\ref{Cauchyprb}) goes beyond the even-wide context of operator difference equations. We indeed emphasize that the treatment does not exclude that the unknowns might depend on some continuous variables and that the operator coefficients $\mathcal{L}_0(n)$
and $\mathcal{L}_1(n)$ may act also upon the single components of $Y_n$. To appreciate the value of this observation consider the following difference-differential equation:
\begin{equation}\label{51}
    Y_{n+2}(t)=\widetilde{\mathcal{L}}_0(n)\dot{Y}_n(t)+\widetilde{\mathcal{L}}_1(n)Y_{n+1}
\end{equation}
where the dot denotes the first time derivative of $Y_n(t)$, which in turn means to derive each component of $Y_n$. The two generally noncommuting operators $\widetilde{\mathcal{L}}_0$ and $\widetilde{\mathcal{L}}_1$ are supposed for simplicity only to linearly mix the component of $Y_n$. Let's then introduce the following operator
\begin{equation}\label{52}
    \mathcal{L}_0(n)=\widetilde{\mathcal{L}}_0(n)\frac{d}{dt};\quad \mathcal{L}_1(n)=\widetilde{\mathcal{L}}_1(n)
\end{equation}
By definition $\widetilde{\mathcal{L}}_0(n)$ and $d/dt$ commute and if $\widetilde{\mathcal{L}}_0(n)$ is represented by a matrix, then
$\mathcal{L}_0(n)$ is a formal matrix as well, whose entries $(\mathcal{L}_0(n))_{ij}$
are differential operators defined as follows:
\begin{equation}\label{53}
    (\mathcal{L}_0(n))_{ij}=[\widetilde{\mathcal{L}}_0(n)]_{ij}\frac{d}{dt}
\end{equation}
where $[\widetilde{\mathcal{L}}_0(n)]_{ij}$ are the $\widetilde{\mathcal{L}}_0(n)$ entries. With the help of this notation, eq. (\ref{51})
formally coincides with eq. (\ref{Cauchyprb}) whose resolutive formula given by eq. (\ref{425}) applies also to difference-differential eq. (\ref{51}),
provided that the initial conditions are appropriately re-interpreted (each component of $\overline{Y}_{0}, \overline{Y}_{1}$ must be thought as functions of $t$).
When $\widetilde{\mathcal{L}}_0(n)$ and $\widetilde{\mathcal{L}}_1(n)$ are defined in accordance with eqs. (\ref{414} )
and (\ref{415}) respectively, where the constant $\rho=1$ is further
adopted, the solution of the Cauchy problem I) ($\overline{Y}_0(t)=0, \overline{Y}_1=\overline{Y}(t)$)
can be cast as follows
\begin{equation}\label{54}Y_n^{I}(t)=
    \left\{\begin{array}{rl}
    M_{-}Y_1(t) & \quad n=2,\\
   D_{+}\frac{d}{dt}+M_{-} & \quad n>2,\mathrm{ even},\\
 D_{+}+M_{-}\frac{d}{dt} &\quad n>2, \mathrm{odd}.
      \end{array}\right.
\end{equation}
As a final remark, we wish to emphasize that the strategic and far-reaching value of the treatment of eq. (\ref{Cauchyprb}) leads to a resolutive formula achieved without requiring a priori the \textquotedblleft mathematical nature\textquotedblright of the unknown \textquotedblleft object\textquotedblright \hspace{0.1cm} $Y_n$.
 Since in the formal framework of a linear second order difference equations eq. (\ref{Cauchyprb}) incorporates  further ingredients such as noncommutativity and is in addition compatible with a possible dependence of $Y_n$ on continuous variables, we believe that out treatment of eq. (\ref{Cauchyprb}) should attract the interest of researchers in many areas from physical mathematics and engineering to economy, biology, social sciences etc, where discrete and/or continuous variables-based mathematical modeling play a central investigational role.

\section*{\large Acknowledgements}
This paper has been written to honor Peter Leach, a deep scientist and a very nice person. A.M. thanks him and the organizers for the pleasant and
warmly atmosphere experienced at Salt Rock, Durban, South Africa, on November 2011 on the occasion of the celebration of his 70'th birthday.
MAJ gratefully acknowledges  the financial support from The Erwin Schrodinger Institute where parts of this work have been done.
The authors thank Professor Andrzej Jamiolkowski for carefully reading the manuscript and for useful comments.

\end{document}